\documentclass{jpp}

\usepackage{amssymb,amsmath,framed}
\usepackage{graphicx}
\usepackage{epstopdf, epsfig}
\usepackage[unicode=true,pdfusetitle,bookmarks=true,bookmarksnumbered=false,bookmarksopen=false,breaklinks=false,pdfborder={0 0 0},backref=false,colorlinks=true]{hyperref}
\hypersetup{citecolor=blue,filecolor=blue,linkcolor=blue,urlcolor=blue}
\usepackage[dvipsnames]{xcolor}

\makeatletter
\@ifpackageloaded{stix} 
{
}{                      
	\usepackage{mathrsfs}
  \usepackage{newtxtext,newtxmath} 	

  \DeclareFontFamily{OML}{txmi1}{}
  \DeclareFontShape{OML}{txmi1}{m}{it}{<->txmi1}{}
  \DeclareSymbolFont{myletter}{OML}{txmi1}{m}{it}
  \DeclareMathSymbol{v}{\mathalpha}{myletter}{`v}

}
\ifx\bm\undefined
  \newcommand{\bm}[1]{{\boldsymbol {\mathrm #1}}} 
\else
  \renewcommand{\bm}[1]{{\boldsymbol {\mathrm #1}}} 
\fi
\makeatother

\newcommand{\f}[2]{\frac{#1}{#2}}
\newcommand{\mr}[1]{\mathrm{#1}}
\newcommand{\lf}{\left}
\newcommand{\ri}{\right}

\newcommand{\dd}[2]{\frac{\rmd#1}{\rmd#2}}
\newcommand{\pp}[2]{\frac{\p #1}{\p #2}}

\newcommand{\Div}{\nabla\cdot}

\newcommand{\nbl}{\nabla}

\newcommand{\+}{{\perp}}
\newcommand{\unit}[1]{\hat{\bm{#1}}}

\makeatletter
\newcommand{\vast}{\bBigg@{4}}
\newcommand{\Vast}{\bBigg@{5}}

\newcommand{\zhat}{\unit{z}} 
 
\newcommand{\yhat}{\unit{y}}

\makeatother

\newcommand{\calI}{\mathcal{I}}

\newcommand{\calN}{\mathcal{N}}
\newcommand{\calO}{\mathcal{O}}

\DeclareMathAlphabet{\mathpzc}{OT1}{pzc}{m}{it}

\newcommand{\rmd}{\mathrm{d}}
\newcommand{\rme}{\mathrm{e}}

\newcommand{\rmi}{\mathrm{i}}

\newcommand{\rmA}{\mathrm{A}}

\newcommand{\rmS}{\mathrm{S}}

\shorttitle{Alfv\'enic vs. compressive fluctuations in MRI turbulence with near-azimuthal mean field}
\shortauthor{Y. Kawazura et al.}

\title{Energy partition between Alfv\'enic and compressive fluctuations in magnetorotational turbulence with near-azimuthal mean magnetic field}

\author{
  Y. Kawazura\aff{1,2,3}\corresp{\email{kawazura@tohoku.ac.jp}},
  A. A. Schekochihin\aff{4,5},
  M. Barnes\aff{4,6}, \\
  W. Dorland\aff{7},
 \and S. A. Balbus\aff{8}
}

\affiliation{
  \aff{1}Frontier Research Institute for Interdisciplinary Sciences, Tohoku University, 6-3 Aoba, Aramaki, Sendai 980-8578, Japan
  \aff{2}Department of Geophysics, Graduate School of Science, Tohoku University, 6-3 Aoba, Aramaki, Aoba-ku, Sendai 980-8578 Japan
  \aff{3}Astrophysical Big Bang Laboratory, RIKEN, 2-1 Hirosawa, Wako, Saitama 351-0198, Japan
  \aff{4}Rudolf Peierls Centre for Theoretical Physics, University of Oxford, Clarendon Laboratory, Parks Road, Oxford OX1 3PU, UK
  \aff{5}Merton College, Oxford OX1 4JD, UK
  \aff{6}University College, Oxford OX1 4AN, UK
  \aff{7}Department of Physics, University of Maryland, College Park, MD 20742-3511, USA
  \aff{8}Oxford Astrophysics, University of Oxford, Denys Wilkinson Building, Keble Road, Oxford OX1 3RH, UK
}

\date{\today}

\begin{document}

\maketitle

\begin{abstract}
The theory of magnetohydrodynamic (MHD) turbulence predicts that Alfv\'enic and slow-mode-like compressive fluctuations are energetically decoupled at small scales in the inertial range.
The partition of energy between these fluctuations determines the nature of dissipation, which, in many astrophysical systems, happens on scales where plasma is collisionless. 
However, when the magnetorotational instability (MRI) drives the turbulence, it is difficult to resolve numerically the scale at which both types of fluctuations start to be decoupled because the MRI energy injection occurs in a broad range of wavenumbers, and both types of fluctuations are usually expected to be coupled even at relatively small scales.
In this study, we focus on collisional MRI turbulence threaded by a near-azimuthal mean magnetic field, which is naturally produced by the differential rotation of a disc.
We show that, in such a case, the decoupling scales are reachable using a reduced MHD model that includes differential-rotation effects.
In our reduced MHD model, the Alfv\'enic and compressive fluctuations are coupled only through the linear terms that are proportional to the angular velocity of the accretion disc.
We numerically solve for the turbulence in this model and show that the Alfv\'enic and compressive fluctuations are decoupled at the small scales of our simulations as the nonlinear energy transfer dominates the linear coupling below the MRI-injection scale.
In the decoupling scales, the energy flux of compressive fluctuations contained in the small scales is almost double that of Alfv\'enic fluctuations.
Finally, we discuss the application of this result to prescriptions of ion-to-electron heating ratio in hot accretion flows.
\end{abstract}

\section{Introduction}
Accretion of matter onto a central massive object is one of the most spectacular astronomical phenomena.
A number of theoretical and numerical studies of accretion flows have been conducted for decades~\citep[see][and references therein]{Balbus1998,Lesur2021}, including the discovery of momentum transport due to turbulence driven by magnetorotational instability~\citep[MRI;][]{Balbus1991}.
On the observational front, the Event Horizon Telescope (EHT) successfully captured an image of a radiating disc at M87~\citep{EHT2019a}, opening the door to direct comparisons between models and observations.
However, there are many unsolved questions in MRI-driven turbulence that are crucial for interpreting such observations. 
In this study, we focus on the energy partition between Alfv\'enic and slow-mode-like compressive fluctuations in collisional MRI turbulence.
This is important in deciding the partition of heating between ions and electrons at dissipation scales, where plasma is collisionless~\citep{Kawazura2020}.

In order to calculate the partition of energy between Alfv\'enic and compressive fluctuations in a numerical simulation, one must access scales small enough that these fluctuations become energetically decoupled. 
It is known that such decoupling is established at the scale where the reduced magnetohydrodynamics (RMHD) approximation, $k_\|/k_\+ \ll 1$ and $|\delta\bm{B}|/B_0 \sim |\delta\bm{u}|/v_\rmA \ll 1$, is satisfied~\citep{Schekochihin2009}.
Here, the subscript $\|$ ($\+$) denotes the component parallel (perpendicular) to the ambient magnetic field, the prefix $\delta$ and the subscript 0 denote the fluctuation and equilibrium fields, respectively, and $v_\rmA$ denotes the Alfv\'en speed.
The RMHD approximation is expected to be satisfied at sufficiently small scales in the inertial range, because the large-scale magnetic field serves as an effective mean field for the fluctuations at the smaller scales~\citep{Kraichnan1965}.
Once the cascade reaches the RMHD range, the partition of Alfv\'enic and compressive fluctuations is maintained down to the ion Larmor scale~\citep{Schekochihin2009}.

While the partition of Alfv\'enic and compressive fluctuations has been studied in externally forced MHD turbulence~\citep{Cho2002,Cho2003,Makwana2020}, none of the previous studies of MRI turbulence have investigated this problem.
Previous MRI turbulence simulations have suggested that, in order to reach the RMHD range, significantly higher numerical resolution is necessary for MRI turbulence than for externally forced MHD turbulence, because there is non-local energy transfer~\citep{Lesur2011}, meaning that the injection range is broad in the Fourier space.

Here, instead of carrying out brute-force high-resolution MHD simulations, we study the partition of Alfv\'enic and compressive fluctuations by reducing the MHD equations to a more tractable form that is valid only when there is a mean magnetic field in approximately azimuthal direction.  
The presence of the near-azimuthal mean magnetic field is a natural consequence of the differential rotation of the disc;
even when the system is initialized with a purely vertical magnetic field, MRI creates a radial magnetic field which will then be twisted in the azimuthal direction.
Indeed, predominantly azimuthal magnetic field is quite often seen both in local and global simulations of MRI turbulence~\citep[e.g.,][]{Suzuki2009,Suzuki2014}.
A statistical analysis of MRI turbulence in incompressible MHD also supports the presence of a near-azimuthal mean field~\citep{Zhdankin2017}.
We show that the RMHD approximation captures the fastest-growing MRI modes in such a system.
We then simulate this type of MRI turbulence numerically and show that the compressive fluctuations carry almost twice as much energy flux as Alfv\'enic fluctuations at the small scales, where the two kinds of fluctuations are decoupled.

\section{Model}
We consider a local shearing-box approximation~\citep{Goldreich1965} for a plasma in Cartesian coordinates $(X,\, Y,\, Z)$ located at a fixed radius $r = r_0$ and rotating with an angular velocity $\bm{\Omega} = \Omega\unit{Z}$, where $X$, $Y$, and $Z$ correspond to the radial, azimuthal, and vertical (rotation-axis) directions.
The MHD equations in these conditions are 
\begin{subequations}
\begin{align}
  &\pp{\rho}{t} + \bm{u}\cdot\nbl\rho + \bm{u}_0\cdot\nbl\rho = -\rho(\Div\bm{u}),
  \label{e:continuity} \\
  &\rho\lf( \pp{}{t} + \bm{u}\cdot\nbl + \bm{u}_0\cdot\nbl \ri)\bm{u} = -\nbl\lf( p + \f{B^2}{8\pi} \ri) + \f{\bm{B}\cdot\nbl\bm{B}}{4\pi} - 2\rho\bm{\Omega}\times\bm{u} - \rho\bm{u}\cdot\nbl\bm{u}_0
  \label{e:e.o.m.}, \\
  &\pp{\bm{B}}{t} + \bm{u}\cdot\nbl\bm{B} + \bm{u}_0\cdot\nbl\bm{B} + \bm{B}(\Div\bm{u}) = \bm{B}\cdot\nbl\bm{u} + \bm{B}\cdot\nbl\bm{u}_0,
  \label{e:induction} \\
  &\pp{p}{t} + \bm{u}\cdot\nbl p + \bm{u}_0\cdot\nbl p + \Gamma p\Div\bm{u} = 0,
  \label{e:adiabatic} 
\end{align}
\end{subequations}
where $\rho$ is the mass density, $\bm{u}$ is the fluid velocity, $\bm{B}$ is the magnetic field, $p$ is the thermal pressure, $\bm{u}_0 \equiv q \bm{X}\times \bm{\Omega}$ is the background shear flow, $q$ is a shear rate, and $\Gamma = 5/3$ is the specific heat ratio.
Hereafter, we only consider Keplerian rotation ($q = 3/2$) and call \eqref{e:continuity}-\eqref{e:adiabatic} the ``full-MHD'' equations.

\begin{figure}
  \begin{center}
    \includegraphics*[width=1.0\textwidth]{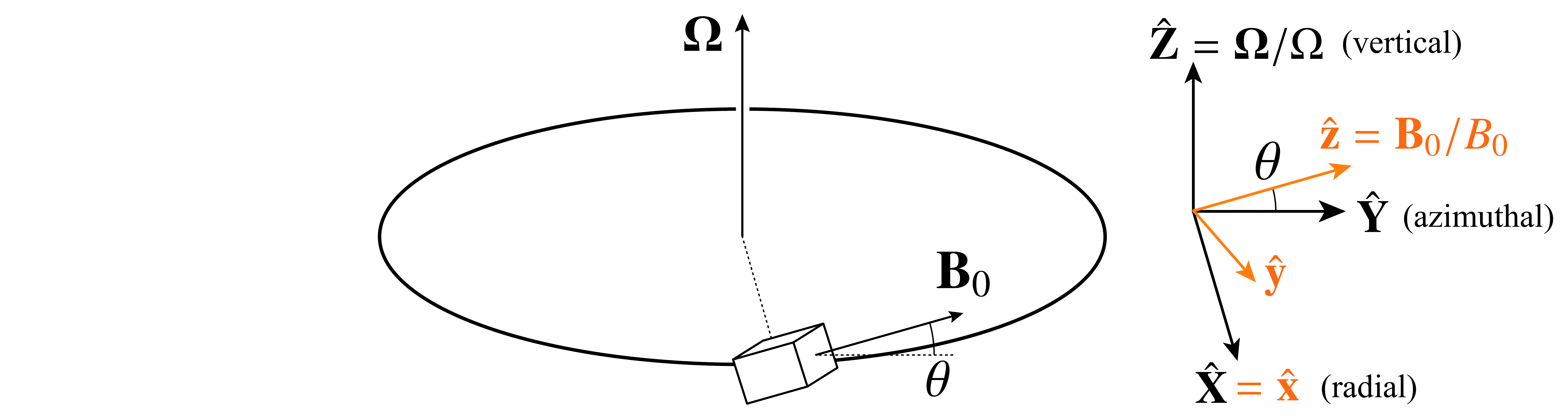}
  \end{center}
  \caption{Schematic of the conventional coordinate system $(X, Y, Z)$ and our tilted coordinate system $(x, y, z)$.}
  \label{f:configuration}
\end{figure}

\cite{Balbus1992a} showed that the fastest-growing MRI modes have $k_Z \to \infty$ when the ambient magnetic field $\bm{B}_0$ approaches the azimuthal direction. 
These fastest-growing modes also satisfy $k_\| v_\rmA/\Omega \simeq 1$.
For a near-azimuthal $\bm{B}_0$, $\hat{\bm{Z}}$ is almost perpendicular to $\bm{B}_0$, meaning that the fastest growing modes satisfy $k_\|/k_\+ \ll 1$.
Therefore, if the fastest-growing modes decide the nature of MRI turbulence at the smaller scales, we can ignore the scales that are outside of the $k_\|/k_\+ \ll 1$ approximation\footnote{
Note that the radial wavenumbers of the nonaxisymmetric modes are time-dependent: $k_X(t) = k_X(0) + q\Omega t k_Y$.
Therefore, these shearing waves inevitably pass the wavenumber domain of $k_x = k_X \gg k_Y \sim k_\|$. 
However, this does not break our approximation $k_\|/k_\+ \ll 1$ because when $\bm{B}_0$ is nearly azimuthal, the fastest-growing modes have large vertical wavenumbers $k_Z \sim k_\+ \gg k_Y$. 
}.
This idea motivates us to impose the RMHD approximation on the full-MHD equations \eqref{e:continuity}-\eqref{e:adiabatic}.
We assume also that the magnetic perturbations are separated from the time-invariant and spatially uniform mean fields as $\bm{B} = \bm{B}_0 + \delta \bm{B}$, where $\bm{B}_0$ is taken to have finite $Y$ (azimuthal) and $Z$ (vertical) components but no $X$ (radial) component.
We assume that the density and pressure are also separated into a constant background and perturbations as $\rho = \rho_0 + \delta\!\rho$ and $p = p_0 + \delta\! p$.
The angle between $\unit{Y}$ and $\bm{B}_0$ is denoted by $\theta \; [0 \le \theta \le \pi; \text{the same definition as \citet{Quataert2002}}]$.
In this study, we focus on the near-azimuthal background field, i.e.,  $\sin\theta \ll 1$.
Then, we introduce a ``tilted'' coordinate set $(x,\, y,\, z)$ in which the $z$-axis is aligned with $\bm{B}_0$, and the $x$-axis is aligned with the $X$-axis (Fig.~\ref{f:configuration}), i.e., $(x,y,z)$ is a rotation of $(X,Y,Z)$ by $\pi/2 - \theta$ about the $\unit{X}$ axis.
When $\sin\theta \ll 1$, $\zhat$ and $\yhat$ almost align with $\hat{\bm{Y}}$ and $-\hat{\bm{Z}}$, respectively.
This tilted coordinate set is more convenient than the standard coordinate set $(X,\, Y,\, Z)$ because $k_z \sim k_\| \ll k_\+$ is a key criterion for the decoupling of Alfv\'enic and compressive fluctuations.
In the standard coordinate set, however, $k_\|$ and $k_\+$ are more difficult to separate, both being a mixture of $k_Y$ and $k_Z$.

Thus, we impose the RMHD ordering $k_\| \ll k_\+,\, \bm{u} \ll v_\rmA$, and $\delta \bm{B} \ll \bm{B}_0$ on \eqref{e:continuity}-\eqref{e:adiabatic}.
We also assume $\p/\p t \sim \Omega \sim k_\| v_\rmA$ and a near-azimuthal $\bm{B}_0$, i.e., $\sin\theta \ll 1$, with $\sin\theta$ the same order of smallness as $k_\|/k_\+$ and $\delta\! B/B_0$.
The assumed anisotropy between $k_\|$ and $k_\+$ is motivated by the critical balance conjecture \citep[CB;][]{Goldreich1995,Goldreich1997}: 
\begin{equation}
  k_\| v_\rmA \sim k_\+ u_\+,
  \label{e:CB}
\end{equation}
which physically means that the time scales of linear wave propagation along $\bm{B}_0$ and nonlinear cascade in the plane perpendicular to $\bm{B}_0$ are of the same order\footnote{In a rapidly rotating fluid, turbulence can also develop anisotropy due to the effect of rotation, leading to $k_Z \ll k_X, k_Y$~\citep[see][and references therein]{Nazarenko2011}. 
However, in our magnetic and differentially rotating system, MRI will inject motions that are in the opposite limit: $k_Z \sim k_\+ \gg k_\| \sim k_Y$ and also $k_Z \gg k_X$ (see Sec.~\ref{s:linear}).
We do not expect the MRI-driven turbulence to be able to access the part of the wavenumber space where the rotational anisotropy is possible.
}.
Then, we obtain the RMHD equations with differential rotation (see appendix~\ref{s:derivation of RMHD} for the detailed derivation): 
\begin{subequations}
\begin{align}
  &\dd{\Psi}{t} = v_\rmA\pp{\Phi}{z},
  \label{e:RMHD Psi} \\
  &\dd{}{t}\nbl_\+^2\Phi = v_\rmA\nbl_\|\nbl_\+^2\Psi - 2\Omega\,\pp{u_\|}{y},
  \label{e:RMHD Phi} \\
  &\dd{u_\|}{t} = v_\rmA^2\nbl_\|\f{\delta\! B_\|}{B_0} + (2 - q)\Omega \,\pp{\Phi}{y},
  \label{e:RMHD upar} \\
  &\dd{}{t}\lf(  1 + \f{v_\rmA^2}{c_\rmS^2} \ri)\f{\delta\! B_\|}{B_0} = \nbl_\| u_\| + \f{q\Omega }{v_\rmA}\,\pp{\Psi}{y},
  \label{e:RMHD Bpar} 
\end{align}
\end{subequations}
where $c_\rmS$ is the sound speed, and $\Phi$ and $\Psi$ are the stream function and magnetic flux function defined by $\bm{u}_\+ = \zhat\times\nbl_\+\Phi$ and $\delta\bm{B}_\+ = \sqrt{4\pi\rho_0}\zhat\times\nbl_\+\Psi$, respectively.
We have also defined $\rmd/\rmd t \equiv \p/\p t + \bm{u}_\perp\cdot\nbl_\perp$ and $\nbl_\| \equiv \p/\p z + (\delta\bm{B}_\+/B_0)\cdot\nbl_\+$.
Hereafter, we call these equations Rotating RMHD (RRMHD)\footnote{Note that \eqref{e:RMHD Psi}-\eqref{e:RMHD Bpar} are akin to the two-dimensional incompressible MHD model~\citep{Julien2006,Morrison2013}, but our model is three-dimensional and applicable to arbitrary $\beta = 8\pi p_0/B_0^2$.}.
When $\Omega = 0$, these become the standard RMHD equations~\citep{Kadomtsev1974,Strauss1976}, which is a long-wavelength limit of gyrokinetics and in which Alfv\'enic and compressive fluctuations are decoupled~\citep{Schekochihin2009}.

One may notice that \eqref{e:RMHD Psi}-\eqref{e:RMHD Bpar} do not have the shearing effect that originates from $\bm{u}_0\cdot\nbl$ terms in \eqref{e:continuity}-\eqref{e:adiabatic}. 
This is due to $k_\|/k_\+ \ll 1$ and $\sin\theta \ll 1$;
in a shearing box, the radial wavenumber depends on time as $k_x(t) = k_x(0) + q\Omega t (k_y\sin\theta + k_\| \cos\theta)$ [see, e.g., the fourth term in the left-hand side of \eqref{e:continuity -2-}], and the time-dependent term on the right-hand side is ordered out because $\sin\theta \sim k_\|/k_x(0) \sim \epsilon$.
However, when we consider a long-time evolution $\Omega t \sim \epsilon^{-1}$, the time dependence is not negligible.
In that case, the non-modal growth of MRI~\citep{Squire2014a,Squire2014b} becomes important.
On the other hand, as we will show below, the eddy turnover time in RRMHD turbulence gets shorter than the disc rotation time, i.e., $k_\+ u_\+/\Omega \gg 1$ immediately below the injection scale (see Fig.~\ref{f:omega_nl}).
Therefore, we do not need to consider a long-time evolution with $\Omega t \sim \epsilon^{-1}$.

As we shall see in the next section, when $\Omega \ne 0$, this system can be MRI unstable.
In the turbulent state, the magnitudes of the nonlinear terms in \eqref{e:RMHD Psi}-\eqref{e:RMHD Bpar} increase as the cascade proceeds to smaller scales, and at some point, the linear terms that are proportional to $\Omega$ become negligible.
Below the scale at which this happens, the turbulence is governed by standard RMHD, and thus Alfv\'enic and compressive fluctuations are decoupled. 
As we will see below, this critical scale roughly corresponds to the scale at which the eddy turnover time becomes shorter than $\Omega^{-1}$.
In other words, when an eddy's lifetime is much shorter than the orbital time of the disc, the effects of the disc's rotation are insignificant.
Therefore, the transient growth effects~\citep{Balbus1992b,Mamatsashvili2013} are absent.
We also note that, with the normalizations $t \Omega \to t$, $z \Omega/v_\rmA \to z$, $x/L_\+ \to x$, $\Phi/L_\+^2\Omega \to \Phi$, $\Psi/L_\+^2\Omega \to \Psi$, $u_\|/L_\+\Omega \to u_\|$, and $v_\rmA \delta\! B_\|/B_0 L_\+\Omega \to \delta\! B_\|$, the rotation rate is no longer a free parameter, and the only remaining parameter is $c_\rmS^2/v_\rmA^2 = \Gamma\beta/2$, where $\beta = 8\pi p_0/B_0^2$.

The nonlinear free-energy invariant of \eqref{e:RMHD Psi}-\eqref{e:RMHD Bpar} consists of Alfv\'enic and compressive portions $W_\mr{tot} = W_\mr{AW} + W_\mr{compr}$, where 
\begin{subequations}
\begin{align}
  W_\mr{AW} &= \f{1}{2}\int\rmd^3\bm{r} \lf[ |\nbl_\+\Phi|^2 + |\nbl_\+\Psi|^2 \ri]
  \label{e:W_AW},\\
  W_\mr{compr} &= \f{1}{2}\int\rmd^3\bm{r} \lf[ u_\|^2 + v_\rmA^2\lf( 1 + \f{v_\rmA^2}{c_\rmS^2} \ri)\f{\delta\! B_\|^2}{B_0^2} \ri].
  \label{e:W_compr}
\end{align}
\end{subequations}
The time evolution of $W_\mr{AW}$ and $W_\mr{compr}$ is given by
\begin{subequations}
\begin{align}
  \dd{W_\mr{AW}}{t} &= - 2\Omega\int\rmd^3\bm{r}\, u_\|\pp{\Phi}{y} \equiv I_\mr{AW},
  \label{e:I_AW}\\
  \dd{W_\mr{compr}}{t} &= q\Omega\int\rmd^3\bm{r}\, \lf[ v_\rmA\f{\delta\! B_\|}{B_0}\pp{\Psi}{y} - u_\|\pp{\Phi}{y} \ri] + 2\Omega\int\rmd^3\bm{r}\, u_\|\pp{\Phi}{y} \equiv I_\mr{compr},
  \label{e:I_compr}
\end{align}
\end{subequations}
where $I_\mr{AW}$ and $I_\mr{compr}$ are the energy injection rates of Alfv\'enic and compressive fluctuations.
Noticing that the second term of $I_\mr{compr}$ is identical to $-I_\mr{AW}$, we may write $I_\mr{compr} = I_\mr{MRI} - I_\mr{AW}$. 
Then the net injection rate by the MRI is $I_\mr{MRI}$, and it goes into compressive fluctuations, which then exchange energy with Alfv\'enic fluctuations at the rate $I_\mr{AW}$ via linear coupling.

\section{Linear MRI of RRMHD}\label{s:linear}
Next, we compare the linear MRI growth rate of full-MHD and RRMHD to show that RRMHD can capture the MRI growth rate of full-MHD when $\bm{B}_0$ is nearly azimuthal, viz., $\sin\theta \ll 1$. 
Here, we focus on the modes that are symmetric with respect to the rotation axis $\unit{Z}$, viz., $k_Y = 0$, which is equivalent to $k_y = -k_z/\tan\theta$.
We focus on these modes because they are the fastest growing ones.
The linear dispersion relation of full-MHD~\citep[Eq. 99]{Balbus1998} is
\begin{multline}
  \lf[\omega^2 - (k_\|v_\rmA)^2\ri]\lf\{ \omega^4 - \lf[k_x^2 + (k_\|/\sin\theta)^2\ri](c_\rmS^2 + v_\rmA^2)\omega^2 + (k_\| v_\rmA)^2\lf[k_x^2 + (k_\|/\sin\theta)^2\ri]c_\rmS^2 \ri\} \\
  = 2\lf\{ (2 - q)\omega^4 - k_\|^2\lf[ 2c_\rmS^2 + (2\cos^2\theta - q)(c_\rmS^2 + v_\rmA^2)/\sin^2\theta \ri]\omega^2 - qc_\rmS^2 v_\rmA^2 k_\|^4/\sin^2\theta \ri\}\Omega^2.
  \label{e:full-MHD dispersion}
\end{multline}
On the other hand, the dispersion relation of RRMHD, \eqref{e:RMHD Psi}-\eqref{e:RMHD Bpar}, is 
\begin{equation}
  \lf[ \omega^2 - (k_\| v_\rmA)^2 \ri]\lf[ \lf(1 + \f{v_\rmA^2}{c_\rmS^2}\ri)\omega^2 - (k_\| v_\rmA)^2 \ri] = 2\lf[ (2 - q)\lf(1 + \f{v_\rmA^2}{c_\rmS^2}\ri)\omega^2 + q(k_\| v_\rmA)^2 \ri]\f{k_y^2}{k_\+^2} \Omega^2.
  \label{e:RMHD dispersion}
\end{equation}
One can show that for both \eqref{e:full-MHD dispersion} and \eqref{e:RMHD dispersion}, $k_x = 0$ gives the fastest-growing mode.
For the RRMHD dispersion relation \eqref{e:RMHD dispersion}, the growth rate does not depend on $k_y$ when $k_x = 0$.
When $\Omega = 0$, \eqref{e:full-MHD dispersion} recovers the Alfv\'en, slow, and fast modes, while \eqref{e:RMHD dispersion} recovers the Alfv\'en and slow modes (the fast mode is eliminated in the RMHD ordering).
The maximum growth rate of RRMHD is given by
\begin{equation}
  \f{\gamma_\mr{max}}{\Omega} = \sqrt{\f{5}{18}\beta\lf[20\beta + 15  - \sqrt{8(50\beta^2 + 75\beta + 18})\ri]},
  \label{e:gamma_max}
\end{equation}
where we have used $q = 3/2$ and $\Gamma = 5/3$.
One finds that $\gamma_\mr{max}$ is an increasing function of $\beta$ ranging from $\gamma_\mr{max}/\Omega \to 0$ for $\beta \to 0$\footnote{For the fastest-growing modes that satisfy $k_X = k_Y = 0$ and $k_\| v_\rmA \sim \Omega$, the locality of the modes in the $Z$ direction, i.e., $k_Z H \gg 1$, implies $\sqrt{\beta}/\sin\theta \gg 1$, where $H = c_\rmS/\Omega$ is the scale height of the disc. While this condition is satisfied in RRMHD as we assume $\beta \sim 1$ and $\sin\theta \ll 1$ in the derivation of RRMHD (appendix~\ref{s:derivation of RMHD}), one must make $\theta$ even smaller in order to make our ordering valid at the low $\beta$ limit.} to $\gamma_\mr{max}/\Omega \to 3/4$ for $\beta \to \infty$.
Note that the high-$\beta$ limit of the maximum growth rate in RRMHD is the same as in full-MHD~\citep[Eq. 114]{Balbus1998}, and the stabilization of MRI at $\beta \to 0$ is consistent with the study by~\citet{Kim2000}, who found that MRI in full-MHD was stabilized when $\beta \to 0$ and $\theta < 30^\circ$.

\begin{figure}
  \begin{center}
    \includegraphics*[width=1.0\textwidth]{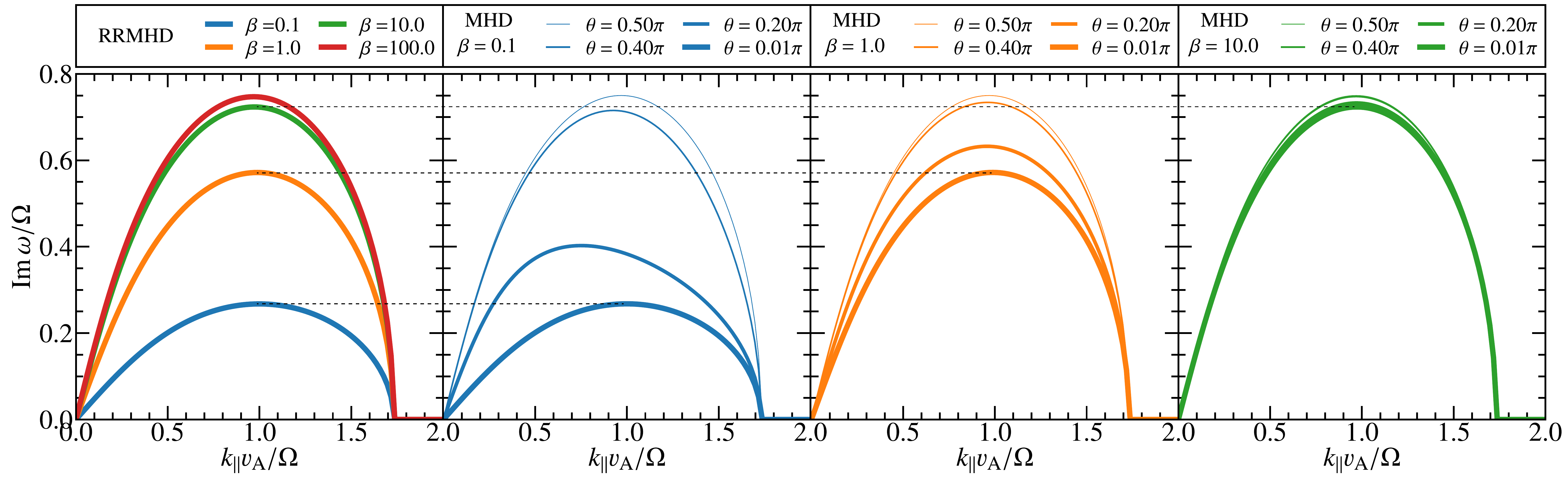}
  \end{center}
  \caption{The linear MRI growth rate of (a) RRMHD and (b)-(d) full-MHD. The line colours correspond to the values of $\beta$ as given in the legend of (a). \hspace{-0.5em} The line thickness for (b)-(d) corresponds to the value of $\theta$ as given in the legends of these panels. The horizontal dotted lines indicate that, independently of $\beta$, the maximum growth rates in RRMHD coincide with those in full-MHD in the limit of $\theta \to 0$.}
  \label{f:dispersion}
\end{figure}

In Fig.~\ref{f:dispersion}, we compare the solutions to \eqref{e:full-MHD dispersion} and \eqref{e:RMHD dispersion}.
Figure~\ref{f:dispersion}(a) shows the growth rate obtained with RRMHD for different values of $\beta$; one finds that $\gamma_\mr{max}$ of RRMHD decreases as $\beta$ decreases as expected from \eqref{e:gamma_max}.
Figures~\ref{f:dispersion}(b)-(d) show the growth rate obtained with full-MHD for different values of $\beta$ and $\theta$.
For full-MHD, the growth rate does not depend on $\beta$ when $\theta = \pi/2$; however, when $\sin\theta \ll 1$, the growth rate decreases as $\beta$ decreases.
Clearly, the growth rates in RRMHD match those in full-MHD with $\sin\theta \ll 1$, meaning that RRMHD captures the fastest-growing MRI modes when $\bm{B}_0$ is nearly azimuthal.

\section{Simulation of MRI turbulence in RRMHD}
Next, we carry out nonlinear simulations of the RRMHD equations to compute the energy partition between the Alfv\'enic and compressive fluctuations in the saturated state of MRI turbulence. 
We solve \eqref{e:RMHD Psi}-\eqref{e:RMHD Bpar} using a 3D pseudo-spectral code \textsc{\textsf{Calliope}}~\citep{Kawazura2022b}.
In order to terminate the energy cascade at small scales, we add hyper-viscous and hyper-resistive terms proportional to $k_\perp^8$ and $k_z^8$ to the right-hand sides of \eqref{e:RMHD Psi}-\eqref{e:RMHD Bpar}.
As mentioned above, the Alfv\'enic and compressive fluctuations are expected to be decoupled below some critical scale where the nonlinear terms start to dominate the linear terms.
We set the computational grids so that this critical scale is well resolved, which we confirm later in this section.
Therefore, the dissipation caused by the hyper-viscosity and hyper-resistivity in \eqref{e:RMHD Psi} and \eqref{e:RMHD Phi} is a measure of the energy flux carried by the Alfv\'enic fluctuations.
Likewise, we can measure the energy flux carried by compressive fluctuations via the hyper-dissipation in \eqref{e:RMHD upar} and \eqref{e:RMHD Bpar}.
We denote the dissipation rates of the Alfv\'enic and compressive fluctuations by $D_\mr{AW}$ and $D_\mr{compr}$, respectively.
The power balance of the system is then
\begin{equation}
  \dd{W_\mr{tot}}{t} = I_\mr{AW} + I_\mr{compr} - D_\mr{AW} - D_\mr{compr}.
\end{equation}
In a statistically stationary state, $\langle I_\mr{AW} \rangle + \langle I_\mr{compr} \rangle = \langle D_\mr{AW} \rangle + \langle D_\mr{compr}\rangle$, where $\langle\cdots\rangle$ denotes the time average.
We set the box size of the simulations as $(L_x, L_y, L_z) = (8\pi L_\+, 2\pi L_\+, 8\pi v_\rmA/\Omega)$ which is discretized by ``low-resolution grids'' $(n_x,\, n_y,\, n_z) = (512,\, 128,\, 512)$, ``medium-resolution grids'' $(n_x,\, n_y,\, n_z) = (1024,\, 256,\, 1024)$, and ``high-resolution grids'' $(n_x,\, n_y,\, n_z) = (1024,\, 256,\, 2048)$.
We choose $L_z$ so that the fastest-growing mode ($k_{z} v_\rmA/\Omega \simeq 1$, as seen in Fig.~\ref{f:dispersion}) fits in the box.
We investigate three cases: $\beta = 0.1, 1$, and $10$.
For all of these values of $\beta$, we start the simulation with the low-resolution grids and run for a sufficiently long time in the nonlinearly saturated state until $\langle \rmd W_\mr{tot}/\rmd t \rangle \simeq 0$ is satisfied before restarting with the higher-resolution grids. 

\begin{figure}
  \begin{center}
    \includegraphics*[width=0.65\textwidth]{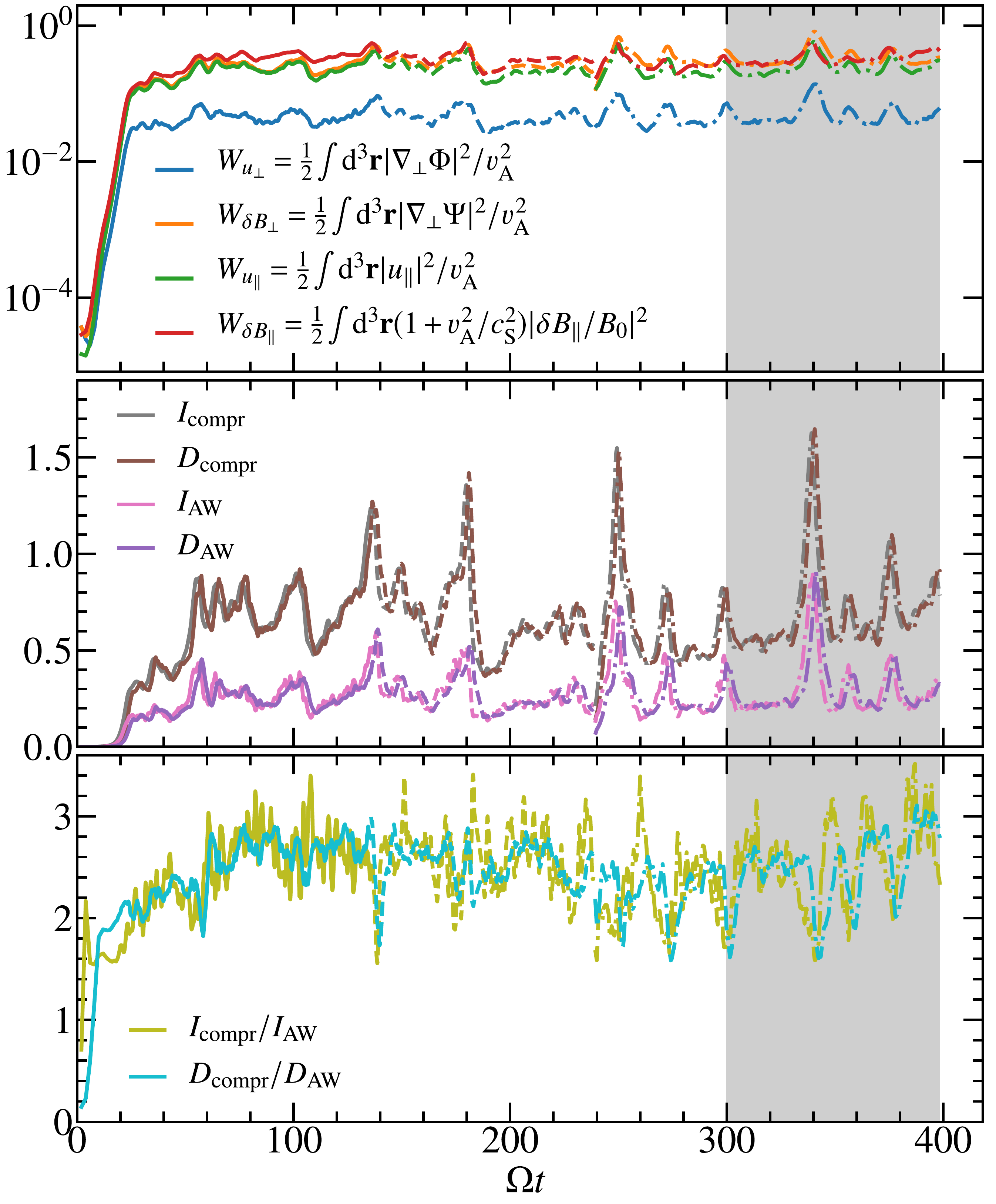}
  \end{center}
  \caption{Time evolution of the $\beta = 1$ run: (top) each component of the free energy \eqref{e:W_AW} and \eqref{e:W_compr} normalized by the total energy averaged over the nonlinearly saturated state, i.e., over the time interval $285\le \Omega t \le 330$; (middle) injection and dissipation rates of Alfv\'enic and compressive fluctuations normalized by the total injection power averaged over the nonlinearly saturated state; (bottom) the compressive-to-Alfv\'enic ratio of injection power $I_\mr{compr}/I_\mr{AW}$ and dissipation rate $D_\mr{compr}/D_\mr{AW}$. The solid, dashed, and dash-dotted lines correspond to the runs with the low-, medium-, and high-resolution grids, respectively. The shaded region indicates the interval used for the time averaging.}
  \label{f:nonlinear t}
\end{figure}
\begin{figure*}
  \begin{center}
    \includegraphics*[width=1.0\textwidth]{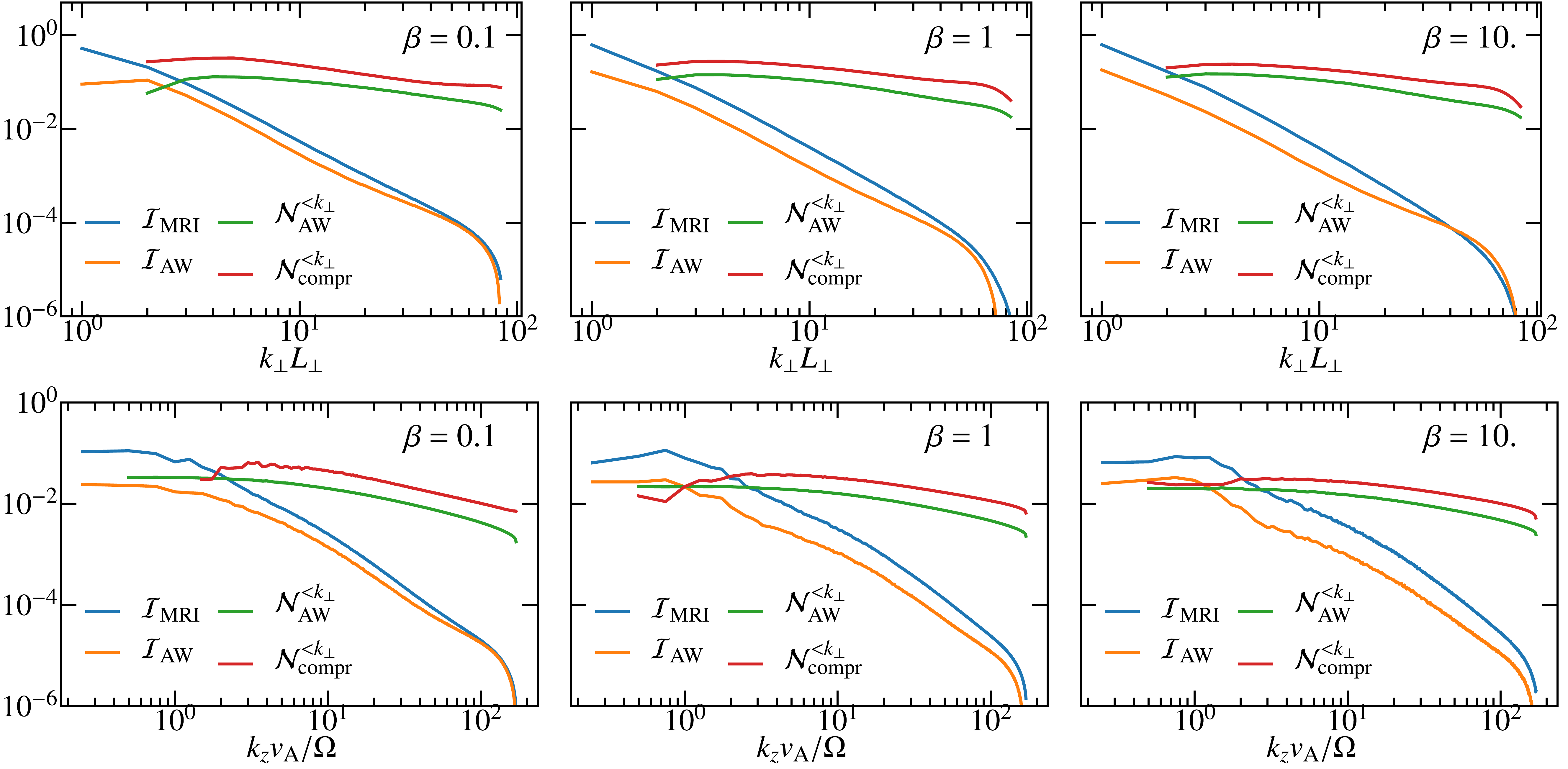}
  \end{center}
  \caption{The spectra of energy injection via MRI, energy exchange between Alfv\'enic and compressive fluctuations, dissipation of Alfv\'enic and compressive fluctuations, and nonlinear transfer vs. (top) $k_\perp$ and (bottom) $k_z$ for (left) $\beta = 0.1$, (middle) $\beta = 1$, and (right) $\beta = 10$. The spectra are normalized by $\langle I_\mr{MRI} \rangle$, integrated over (top) $k_z$ and (bottom) $k_\+$, and averaged over the time interval shown by the shaded area in Fig.~\ref{f:nonlinear t}.}
  \label{f:coupling vs transfer}
\end{figure*}
\begin{figure}
  \begin{center}
    \includegraphics*[width=0.55\textwidth]{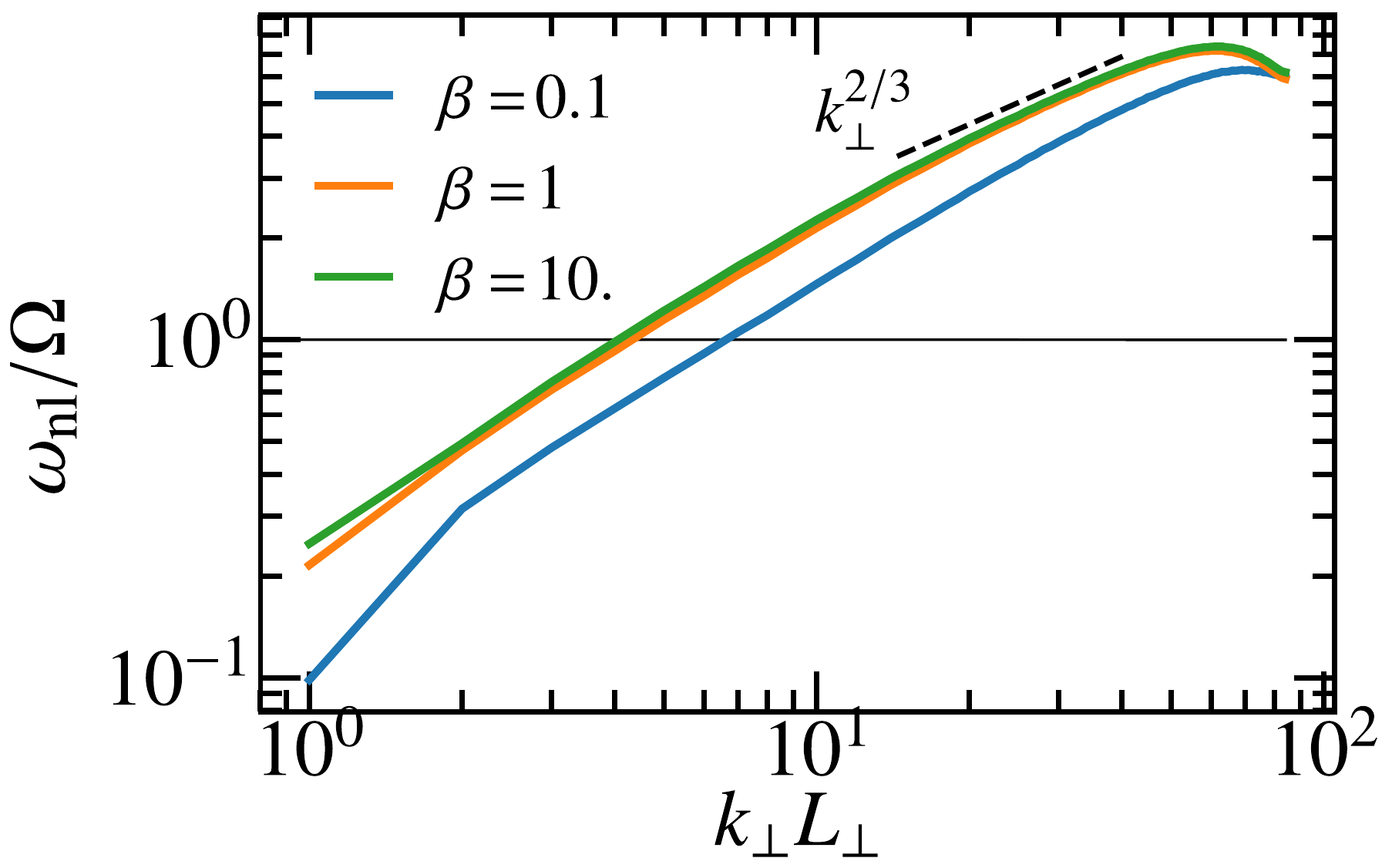}
  \end{center}
  \caption{Normalized eddy turnover frequency $\omega_\mr{nl}/(\Omega\cos\theta)$ vs. $k_\+ L_\+$ averaged over the time interval shown by the shaded area in Fig.~\ref{f:nonlinear t} for higher-resolution runs. The effect of differential rotation is negligible where the value is greater than unity. 
  }
  \label{f:omega_nl}
\end{figure}

Figure~\ref{f:nonlinear t} shows the time evolution of the free energy ($W_\mr{AW}$ and $W_\mr{compr}$), the power balance ($I_\mr{AW}$, $I_\mr{compr}$, $D_\mr{AW}$, $D_\mr{compr}$), the compressive-to-Alfv\'enic energy-injection ratio $I_\mr{compr}/I_\mr{AW}$, and the dissipation ratio $D_\mr{compr}/D_\mr{AW}$. 
From the top panel, one finds that the linear-growth phase occurs at $10 \lesssim \Omega t \lesssim 20$ and is followed by the nonlinearly saturated turbulent phase.
While the Alfv\'enic energy consists predominantly of $\delta\! B_\+$, the compressive energy has almost the same contribution from $u_\|$ and $\delta\! B_\|$. 
We have confirmed that this trend is the same for $\beta = 0.1$ and $10$.
The middle panel shows that the energy injection balances with the energy dissipation.
Interestingly, the amount of Alfv\'enic injection $I_\mr{AW}$ balances with Alfv\'enic dissipation $D_\mr{AW}$, and likewise, the compressive injection $I_\mr{compr}$ and dissipation $D_\mr{compr}$ are in balance. 
So, in the saturated state, there is, on average, barely any net nonlinear energy exchange between the two components of the turbulence -- even at larger scales, where they are not formally decoupled.
We have confirmed that this is also the case for $\beta = 0.1$ and $10$.
As we will see later, this may be due to the fact that the critical scale at which the Alfv\'enic and compressive fluctuations decouple is located close to the injection scale~(see Figs.~\ref{f:coupling vs transfer} and \ref{f:omega_nl}).
The bottom panel shows the evolution of $I_\mr{compr}/I_\mr{AW}$ and $D_\mr{compr}/D_\mr{AW}$.
One finds that both ratios are $\simeq 2-2.5$ in the nonlinear state. 
These values are almost the same for the runs with low-resolution grids (solid lines), medium-resolution grids (dashed lines), and high-resolution grids (dash-dotted lines).

\begin{figure}
  \begin{center}
    \includegraphics*[width=0.65\textwidth]{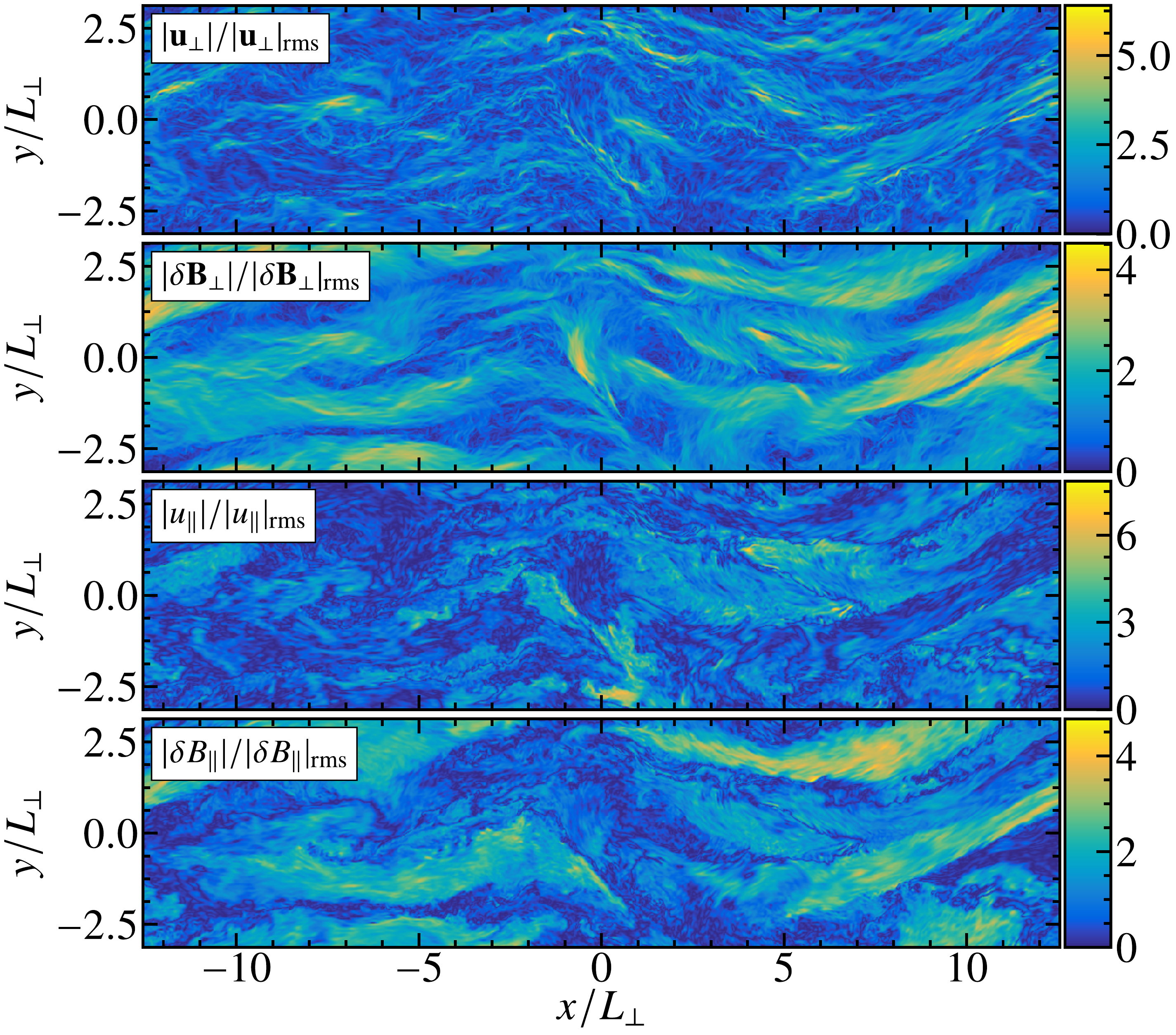}
  \end{center}
  \caption{Snapshots of (from top to bottom) $|\bm{u}_\+|$, $|\delta\bm{B}_\+|$, $|u_\||$, and $|\delta\! B_\||$, each normalized by its own rms value. These snapshots are taken at $\Omega t = 395$, $z = 0$, and for $\beta = 1$.}
  \label{f:nonlinear fields}
\end{figure}

Figure~\ref{f:nonlinear fields} shows snapshots of the turbulent fields. 
Structures are elongated in the $x$ direction, corresponding to the remnants of ``channel flows'' driven by MRI, also seen in other shearing-box simulations of full-MHD~\citep[e.g.,][]{Hawley1992,Hirai2018}.
Note that our $\yhat$ direction is almost vertical within the accretion disc, $\yhat \simeq -\bm{\Omega}/\Omega$ (see Fig.~\ref{f:configuration}).
For the Alfv\'enic fields, one finds that $\bm{u}_\+$ has smaller-scale filamentary structures than $\delta\bm{B}_\+$.
In contrast, for the compressive fields, the level of filamentation is the same between $u_\|$ and $\delta\! B_\|$.
We have found this tendency also for the $\beta = 0.1$ and 10 cases.

\begin{figure*}
  \begin{center}
    \includegraphics*[width=1.0\textwidth]{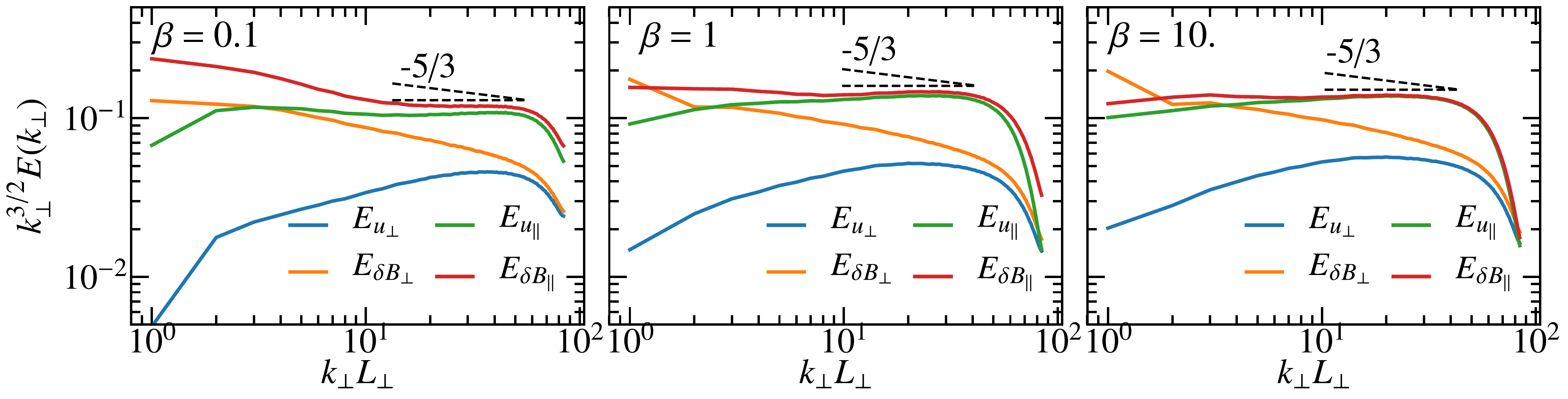}
  \end{center}
  \caption{Magnetic and kinetic spectra compensated by $k_\+^{3/2}$ and averaged over the time interval shown by the shaded area in Fig.~\ref{f:nonlinear t} for high-resolution runs with (left) $\beta = 0.1$, (middle) $\beta = 1$, and (right) $\beta = 10$. The dashed lines indicate the -3/2 and -5/3 slopes.}
  \label{f:nonlinear kperp}
\end{figure*}

\begin{figure}
  \begin{center}
    \includegraphics*[width=0.6\textwidth]{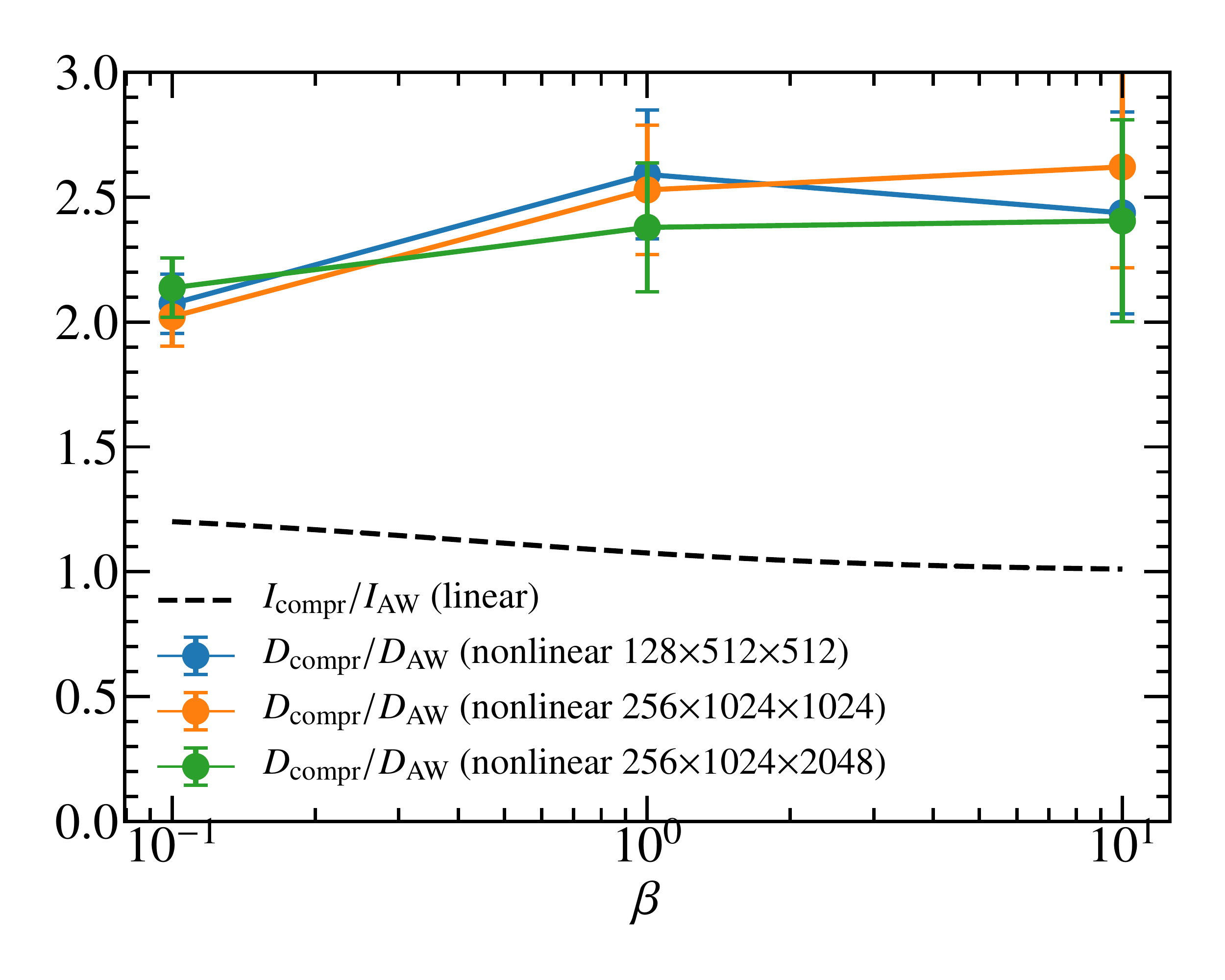}
  \end{center}
  \caption{Partition of energy flux between Alfv\'enic and compressive fluctuations vs. $\beta$: (black) $I_\mr{compr}/I_\mr{AW}$ calculated by the eigenfunctions of linear dispersion relation \eqref{e:RMHD dispersion} and $D_\mr{compr}/D_\mr{AW}$ calculated by the nonlinear simulation with (blue) low-resolution grids, (orange) medium-resolution grids, (green) high-resolution grids. Error bars for the nonlinear simulations are estimated by calculating the standard deviation over the averaging interval.}
  \label{f:partition vs beta}
\end{figure}

The difference of filamentation levels is more transparent in Fig.~\ref{f:nonlinear kperp}, which shows the energy spectra of all fields vs. $k_\+$, compensated by $k_\+^{3/2}$.
Here, the energy spectrum of each integrand in \eqref{e:W_AW} and \eqref{e:W_compr} is denoted by $E$ with the corresponding subscript.
We find that, for the compressive fields, both $E_{u_\|}$ and $E_{\delta\! B_\|}$ have $\simeq -3/2$ slope, while the slopes of the Alfv\'enic fields are not identifiable with the current numerical resolution\footnote{Currently, -3/2 spectral slope is considered to be more likely for the Alfv\'enic cascade based on theoretical arguments and observational evidence~\citep[see][and references therein]{Schekochihin2020}.}.
Independently of $\beta$, $E_{u_\+}$ is subdominant compared to $E_{\delta\! B_\+}$ at the injection scales, whereas $E_{u_\|}$ and $E_{\delta\! B_\|}$ have almost the same amplitudes throughout the whole $k_\+$ range.
It is well known that full-MHD simulations of MRI turbulence yield magnetically dominated spectra at large scales~\citep[e.g.,][]{Lesur2011,Walker2016,Kimura2016,Sun2021}, due to generation of azimuthal magnetic field through the shear-flow effect.
However, this mechanism cannot explain $E_{\delta\! B_\+} \gg E_{u_\+}$ in RRMHD because the shear flow does not directly produce $\delta \bm{B}_\+$, as one can see in \eqref{e:RMHD Psi}. 
Instead we can explain the dominance of $E_{\delta\! B_\+}$ by the linear relation $\Psi/\Phi = k_\| v_\rmA/\gamma$ given by \eqref{e:RMHD Psi}, where $\gamma$ is the growth rate of MRI.
For the fastest growing mode, $k_\| v_\rmA/\Omega \simeq 1$ and $\gamma/\Omega < 1$ (see Fig.~\ref{f:dispersion}), meaning that the linear MRI in RRMHD excites $\delta\! B_\+$ preferentially over $u_\+$.
One also finds from Fig.~\ref{f:nonlinear kperp} that the disparity between $\delta\! B_\+$ and $u_\+$ gets smaller as $\beta$ increases. 
More specifically, at $k_\+ L_\+ = 1$, $E_{\delta\! B_\+}/E_{u_\+} \simeq$ 27, 12, and 10 for $\beta = $ 0.1, 1, and 10, respectively, being consistent with the fact that $\gamma_\mr{max}/\Omega$ is an increasing function of $\beta$. 
Nonetheless, the absolute values of the ratio are somewhat different from the linear estimate $(\Psi/\Phi)^2 \simeq$ 14, 3 and 2 for $\beta = $ 0.1, 1, and 10, respectively, for the fastest-growing mode.
This indicates that the nonlinear effect is important, and, indeed, as we will see in Fig.~\ref{f:partition vs beta}, the partition of energy flux between Alfv\'enic and compressive fluctuations is different between the linear calculation and nonlinear simulations. 

It is worthwhile to compare our spectra with the incompressible MRI simulation by~\citet{Walker2016}, which is the highest-resolution shearing box turbulence to date.
They found that the slope of the magnetic field spectrum was close to -3/2 when the azimuthal component $B_Y$ was subtracted.
They also found nearly -3/2 spectral slope for the velocity field as well.
These spectra bear a resemblance to our spectra (Fig.~\ref{f:nonlinear kperp}).
Note, however, that $B_Y$ is not necessarily the true mean magnetic field $\bm{B}_0$, and thus, their magnetic spectrum $B_X^2 + B_Z^2$ is presumably a mixture of parallel and perpendicular fluctuations.

In order to investigate the decoupling of Alfv\'enic and compressive fluctuations, we compare the spectra of energy injection via MRI ($I_\mr{MRI}$), the energy exchange between the Alfv\'enic and compressive fluctuations ($I_\mr{AW}$), and the nonlinear energy transfer, defined below.
Since the coupling between the Alfv\'enic and compressive fluctuations exists only through the linear terms, the two types of fluctuations are decoupled when the nonlinear energy transfer overwhelms $I_\mr{AW}$.
The nonlinear energy transfer from all modes with wavenumber magnitudes smaller than $k_\+$ are defined by~\citep{Alexakis2005,Grete2017,St-Onge2020}
\begin{align}
  \calN_\mr{AW}^{<k_\+} &\equiv \sum_{|q_\+| = k_\+ }\bigg[ -\bm{u}_{\+\bm{q}}^*\cdot\lf(\bm{u}_\+\cdot\nbl_\+\bm{u}_\+^{[<k_\+]}\ri)_\bm{q} + \bm{u}_{\+\bm{q}}^*\cdot\lf(\bm{b}_\+\cdot\nbl_\+\bm{b}_\+^{[<k_\+]}\ri)_\bm{q} \\
  & \hspace{4.45em}- \bm{b}_{\+\bm{q}}^*\cdot\lf(\bm{u}_\+\cdot\nbl_\+\bm{b}_\+^{[<k_\+]}\ri)_\bm{q} + \bm{b}_{\+\bm{q}}^*\cdot\lf(\bm{b}_\+\cdot\nbl_\+\bm{u}_\+^{[<k_\+]}\ri)_\bm{q} \bigg]\\
  \calN_\mr{compr}^{<k_\+} &\equiv \sum_{|q_\+| = k_\+ }\bigg[ -u_{\|\bm{q}}^*\cdot\lf(\bm{u}_\+\cdot\nbl_\+ u_\|^{[<k_\+]}\ri)_\bm{q} + u_{\|\bm{q}}^*\cdot\lf(\bm{b}_\+\cdot\nbl_\+ b_\|^{[<k_\+]}\ri)_\bm{q} \\
  & \hspace{4.45em}- b_{\|\bm{q}}^*\cdot\lf(\bm{u}_\+\cdot\nbl_\+ b_\|^{[<k_\+]}\ri)_\bm{q} + b_{\|\bm{q}}^*\cdot\lf(\bm{b}_\+\cdot\nbl_\+ u_\|^{[<k_\+]}\ri)_\bm{q} \bigg],
\end{align}
where $f^{[<k_\+]}(\bm{r}) \equiv \sum_{q_z}\sum_{q_\+ < k_\+} f_\bm{q}\rme^{\rmi\bm{q}\cdot\bm{r}}$, $\bm{b} \equiv v_\rmA \delta \bm{B}/B_0$.
We also define the spectra of MRI injection $I_\mr{MRI} = \sum_\bm{k}\calI_\mr{MRI}(\bm{k})$ and energy exchange $I_\mr{AW} = \sum_\bm{k}\calI_\mr{AW}(\bm{k})$.
The top panels of Fig.~\ref{f:coupling vs transfer} show the perpendicular spectra of injection, exchange, and nonlinear energy transfer.
Both the injection and exchange peak near the box scale and drop quickly at smaller scales, while the nonlinear energy transfer is relatively constant throughout the $k_\+$-range.
Consequently, the nonlinear energy transfer overwhelms the injection and coupling immediately below the box scale.
The bottom panels of Fig.~\ref{f:coupling vs transfer} show the $z$ spectra of the same quantities.
The peak of the injection is located around $k_z v_\rmA/\Omega \simeq 1$, and thus, the injection scale corresponds to the fastest-growing modes.
In the same way as the perpendicular spectra, the injection and exchange drop quickly at scales smaller than that of the fastest-growing mode and are overwhelmed by the nonlinear energy transfer.
Therefore, in the small scales of our simulations, the coupling between Alfv\'enic and compressive fluctuations is negligible. 

While the spectral comparison shown in Fig.~\ref{f:coupling vs transfer} is the most direct proof of the decoupling of Alfv\'enic and compressive fluctuations, we expect that the ratio between the eddy-turnover rate and the angular velocity of the accretion disc can also be a proxy for the measurement of the decoupling\footnote{
Note that~\cite{Walker2016} used a quantity similar to $\omega_\mr{nl}/\Omega$ to identify the energy-injection range in incompressible MHD simulations. 
More specifically, they found that the outer scale $\lambda_0$ and the spatial average of turbulence intensity $v_0$ satisfy $v_0/\lambda_0 \sim \rmd\Omega/\rmd\!\ln r$, where $\rmd\Omega/\rmd\!\ln r$ is the local shear rate. 
Since full-MHD has a coupling between the Alfv\'enic and compressive fluctuations through the nonlinear terms, $\omega_\mr{nl}/\Omega$ cannot be used formally as a measurement of decoupling between the two types of fluctuations.
In RRMHD, on the other hand, the decoupling is guaranteed when $\omega_\mr{nl}/\Omega \gg 1$ as demonstrated in Fig.~\ref{f:coupling vs transfer} and \ref{f:omega_nl}.
As an accretion disc tends to produce near-azimuthal mean field, we expect that $\omega_\mr{nl}/\Omega \gg 1$ can still be a proxy for the measurement of the decoupling.
}.
In Fig.~\ref{f:omega_nl}, we plot the eddy-turnover rate $\omega_\mr{nl} \sim|\bm{u}_\+\cdot\nbl_\+| \sim k_\+ u_\+ \sim k_\+^{3/2}E_{u_\+}^{1/2}$ normalized by $\Omega$.
One finds that this value is an increasing function of $k_\+ L_\+$ and exceeds unity at some scale. 
As mentioned above, when $\omega_\mr{nl}/\Omega$ is much larger than unity, the effect of differential rotation is expected to be negligible, and the turbulence obeys the standard RMHD where the Alfv\'enic and compressive fluctuations are decoupled~\citep{Schekochihin2009}.
The scale at which $\omega_\mr{nl}/\Omega \simeq 1$ is not much smaller than the injection scale, which is consistent with the fact that the nonlinear energy transfer overwhelms the MRI injection immediately below the injection scale, as shown in Fig.~\ref{f:coupling vs transfer}.
In general, $\omega_\mr{nl}/\Omega$ is easier to use as an indicator of decoupling because computing the nonlinear energy transfer is numerically cumbersome.

As the decoupling of Alfv\'enic and compressive fluctuations in our simulations has been demonstrated in Figs.~\ref{f:coupling vs transfer} and \ref{f:omega_nl}, we now calculate the partition of energy flux carried by these two types of fluctuations. 
Figure~\ref{f:partition vs beta} shows the dependence of $\langle D_\mr{compr} \rangle/\langle D_\mr{AW} \rangle$ on $\beta$.
We find that $\langle D_\mr{compr} \rangle/\langle D_\mr{AW} \rangle$ is between 2 and 2.5 for all values of $\beta$ that we studied, without an obvious trend.
Since Alfv\'enic and compressive fluctuations are decoupled, the result in Fig~\ref{f:partition vs beta} would not be changed for a finer-resolution simulation.
Indeed, we have found almost identical values of $\langle D_\mr{compr} \rangle/\langle D_\mr{AW} \rangle$ in our simulations conducted at all resolutions, from low to high.
This is because, as seen in Fig.~\ref{f:coupling vs transfer}, even the low-resolution-grid runs resolve the critical scale where the nonlinear energy transfer dominates the linear coupling.  

Note that the values of $\langle D_\mr{compr} \rangle/\langle D_\mr{AW} \rangle$ obtained from our nonlinear simulations are different from the values of $I_\mr{compr}/I_\mr{AW}$ [see \eqref{e:I_compr/I_AW} for the definition] computed ``quasilinearly'' for the fastest-growing linear MRI modes (black dashed line in Fig~\ref{f:partition vs beta}), the latter value being close to unity. 
This indicates that, even though the decoupling of Alfv\'enic and compressive fluctuations starts relatively near the injection scale, the preferential excitation of compressive fluctuations in MRI turbulence is the consequence of nonlinear effects, i.e., of the way in which the nonlinearity removes the energy injected by the MRI from the injection scale and transfers it into the two turbulent cascade.

\section{Application to ion-to-electron heating prescription in hot accretion flows}
In this section, we discuss the application of our findings to hot accretion flows, such as M87 and Sgr A*, together with some important caveats. 
Numerical simulations of gyrokinetic turbulence have revealed that the partition between ion and electron heating is crucially sensitive to the compressive-to-Alfv\'enic injection power ratio $P_\mr{compr}/P_\mr{AW}$ at the ion Larmor scale~\citep{Kawazura2020}\footnote{Particle-in-cell simulations of relativistic turbulence have found a similar dependence of ion-to-electron heating ratio on the compressibility of energy injection~\citep{Zhdankin2021}.}. 
Since the compressive and Alfv\'enic energy fluxes computed in our simulations are supposed to cascade down to the ion Larmor scale independently, it is straightforward to infer that $P_\mr{compr}/P_\mr{AW}$ is equal to $D_\mr{compr}/D_\mr{AW} \simeq 2-2.5$, as we found in Fig.~\ref{f:partition vs beta}.
Therefore, we can combine the results of this paper with our previous stydy of gyrokinetic turbulence to formulate the ion-to-electron heating prescription that incorporates both driving of turbulence via MRI at MHD scales and the dissipation at kinetic scales. 
Substituting $P_\mr{compr}/P_\mr{AW} = 2$ in (14) of ~\cite{Kawazura2019}, one obtains
\begin{equation}
  \f{Q_\rmi}{Q_\rme}(\beta_\rmi,\, T_\rmi/T_\rme) = \f{35}{1 + (\beta_\rmi/15)^{-1.4} \rme^{-0.1/(T_\rmi/T_\rme)}} + 2,
  \label{e:prescription}
\end{equation}
where $T_\rmi/T_\rme$ is the ion-to-electron background temperature ratio, and $\beta_\rmi$ is the ion beta.

This prescription is a step forward from the currently used heating prescription and may help improve the quality of hot accretion flow modelling~\citep[e.g.,][]{Chael2018,Chael2019}.
However, one must bear in mind that a number of heating channels are missing in~\eqref{e:prescription}.  
First, we do not consider spiral density waves~\citep{Heinemann2009} which are outside the RMHD ordering as they have no vertical structure, i.e., $k_Z \simeq 0$.
The excitation of these waves may change the partition between Alfv\'enic and compressive fluctuations. 
Note that these waves form weak shocks and dissipate into thermal energy, but the amount of heating due to this dissipation is very little. 

Second, while we have only considered collisional MRI in this study, the mean free path of hot accretion flows is almost equal to, or longer than, the scale height of the disc, meaning that MRI is supposed to be collisionless\footnote{Nonetheless, most of the extant general-relativistic global simulations have solved collisional MHD, except for only a few studies using general relativistic Braginskii model~\citep{Chandra2015,Foucart2016,Chandra2017,Foucart2017} that takes into account weakly collisional effects.}.
When MRI is collisionless, the viscous stress due to pressure anisotropy gives rise to a new heating channel.
About 50\% of total injected power may be directly converted into heat at large scales by this viscous stress, which would not cascade down to the ion Larmor scale~\citep{Sharma2007,Kempski2019}\footnote{In contrast to heating, the characteristics of turbulence such as the nonlinear saturation level and angular momentum transport are almost the same between collisional MRI and collisionless one~\citep{Sharma2006,Kunz2016,Foucart2016,Squire2017,Kempski2019}.}.

Third, even if these additional heating channels at large scales are absent, there are other heating channels at the ion Larmor scale that are not captured by standard gyrokinetics~\citep[see Sec.~II A in ][for a detailed discussion]{Kawazura2020}, e.g., cyclotron heating~\citep{Cranmer1999}, stochastic heating~\citep{Chandran2010}, and background pressure anisotropy~\citep{Kunz2018}.

Thus, our heating prescription~\eqref{e:prescription} is only the simplest possible model that considers both MRI injection and kinetic dissipation. 
Including the missing heating channels is an important task for future work.

\section{Conclusions}
In this study, we have calculated the energy partition between Alfv\'enic and compressive fluctuations in turbulence driven by MRI with near-azimuthal mean magnetic fields.
The fastest-growing MRI modes are correctly captured by RMHD with differential rotation (RRMHD) because they satisfy $|k_Z/k_Y| \sim |k_\+/k_\|| \gg 1$ when the background field is nearly azimuthal~\citep{Balbus1992b}.
In RRMHD, the Alfv\'enic and compressive fluctuations are coupled only through the linear terms that are proportional to the angular velocity of the accretion disc.
We have carried out nonlinear simulations of RRMHD and showed that the nonlinear energy transfer overwhelms the linear coupling immediately below the injection scale.
Thus, the two kinds of fluctuations are decoupled at the small scales in our simulations.
This is because, below the injection scale, the eddy turnover time is much shorter than the disc rotation time, i.e., $\omega_\mr{nl}/\Omega \gg 1$.
Most importantly, the energy flux carried by the compressive fluctuations is more than double that carried by the Alfv\'enic fluctuations at the decoupled scales --- a result reflecting the interaction between MRI injection and nonlinearity at the injection scale and distinct from a ``quasilinear'' estimate (which suggests near equipartition).

While these findings suggest that RRMHD is a useful model for studying MRI turbulence, we would like to stress the following two limitations of the RMHD approach for MRI-driven turbulence in accretion flows. 
First, we assume a near-azimuthal constant mean magnetic field.
This may be quite restrictive: e.g., global MHD simulations~\citep[e.g.,][]{Suzuki2014} sometimes exhibit non-azimuthal components of magnetic field.
Secondly, we assume that $k_\|/k_\+ \ll 1$ is already satisfied at a larger scale than the critical scale where $\omega_\mr{nl}/\Omega \sim 1$.
If this were not to hold, the rotation effects in full-MHD may become negligible at scales larger than those where the RMHD approximation is already satisfied, and our RRMHD model would not be a good model of MRI turbulence at the decoupling scale.
In such a case, the turbulence in the RMHD range would not be driven by MRI, but by the cascade from the full-MHD scales.
A simulation of full-MHD with extreme resolutions is necessary to explore this possibility.

\section*{Acknowledgements}
YK thanks M.~Kunz for fruitful discussions.
YK, AAS, MB, and SAB were supported in part by the STFC grant ST/N000919/1;
the work of AAS and MB was also supported in part by the EPSRC grant EP/R034737/1.
YK was supported by JSPS KAKENHI grants JP19K23451 and JP20K14509. 
Numerical computations reported here were carried out on Cray XC50 at Center for Computational Astrophysics in National Astronomical Observatory of Japan, on the computing resource at Kyushu University, on Oakforest-PACS and Oakbridge-CX at the University of Tokyo, and on Flow at Nagoya University.

\section*{Declaration of interests}
The authors report no conflict of interest.

\appendix
\section{Derivation of RRMHD model}\label{s:derivation of RMHD}
Here, we explicitly derive \eqref{e:RMHD Psi}-\eqref{e:RMHD Bpar} from \eqref{e:continuity}-\eqref{e:adiabatic}.
The way we do it is mostly the same as the derivation of (17), (18), (25), and (26) in \citet{Schekochihin2009}, but with account taken of the differential rotation of the disc.
We start by considering the following ordering:
\begin{align}
  \f{\bm{u}}{v_\rmA} \sim \f{\delta\bm{B}}{B_0} \sim \f{k_\|}{k_\+} \sim \sin\theta \sim \epsilon, \quad \pp{}{t} \sim \Omega \sim k_\| v_\rmA \equiv \omega.
\end{align}
Then, the order of each term in \eqref{e:continuity} is estimated as follows:
\begin{multline}
  \underbrace{\pp{}{t}\f{\delta\! \rho}{\rho_0}}_{\epsilon^1\omega} + \underbrace{u_\|\pp{}{z}\f{\delta\! \rho}{\rho_0}}_{\epsilon^2\omega} + \underbrace{\bm{u}_\+\cdot\nbl_\+\f{\delta\! \rho}{\rho_0}}_{\epsilon^1\omega} - \underbrace{q\Omega x\sin\theta\Bigg(\pp{}{y} + \f{1}{\tan\theta}\pp{}{z}\Bigg)}_{\epsilon^2\omega}\f{\delta\! \rho}{\rho_0} \\
  = -\Bigg( \underbrace{\pp{u_\|}{z}}_{\epsilon^1\omega} + \underbrace{\nbl_\+\cdot\bm{u}_\+}_{\epsilon^0\omega} \Bigg) - \f{\delta\! \rho}{\rho_0}\Bigg( \underbrace{\pp{u_\|}{z}}_{\epsilon^2\omega} + \underbrace{\nbl_\+\cdot\bm{u}_\+}_{\epsilon^1\omega} \Bigg).
  \label{e:continuity -2-}
\end{multline}
To order $\calO(\epsilon^0\omega)$, we obtain $\nbl_\+\cdot\bm{u}_\+ = 0$.
Likewise, to lowest-order, $\Div \delta\bm{B} = 0$ gives $\nbl_\+\cdot \delta\bm{B}_\+ = 0$.
Therefore, we may write $\bm{u}_\+$ and $\delta \bm{B}_\+$ in terms of stream and flux functions: 
\begin{equation}
  \bm{u}_\+ = \zhat\times\nbl_\+\Phi, \quad \f{\delta\bm{B}_\+}{B_0} = \f{\zhat\times\nbl_\+\Psi}{v_\rmA}.
\end{equation}
Then, the $\calO(\epsilon^1\omega)$ terms in \eqref{e:continuity -2-} yield
\begin{equation}
  \lf( \pp{}{t} + \bm{u}_\+\cdot\nbl_\+ \ri)\f{\delta\! \rho}{\rho_0} = -\pp{u_\|}{z}.
  \label{e:continuity O(1)}
\end{equation}
Note that the shearing term, viz., the fourth term in the left-hand side of \eqref{e:continuity -2-}, is ordered out. 
As we will show shortly, the shearing terms in other equations are also ordered out.

Under the same ordering, terms in \eqref{e:e.o.m.} are ordered as follows:
\begin{multline}
  \underbrace{\pp{\bm{u}}{t}}_{\epsilon^1\omega v_\rmA} + \underbrace{u_\|\pp{\bm{u}}{z}}_{\epsilon^2\omega v_\rmA} + \underbrace{\bm{u}_\+\cdot\nbl_\+\bm{u}}_{\epsilon^1\omega v_\rmA} - \underbrace{q\Omega x\sin\theta\Bigg(\pp{}{y} +  \f{1}{\tan\theta}\pp{}{z}\Bigg)}_{\epsilon^2\omega v_\rmA}\bm{u} = -\zhat\pp{}{z}\Bigg[ \underbrace{\f{c_\rmS^2}{\Gamma} \f{\delta\! p}{p_0}}_{\epsilon^1\omega v_\rmA} + v_\rmA^2\Bigg( \underbrace{\f{1}{2}\f{|\delta\bm{B}|^2}{B_0^2}}_{\epsilon^2\omega v_\rmA} + \underbrace{\f{\delta\! B_\|}{B_0}}_{\epsilon^1\omega v_\rmA} \Bigg) \Bigg]\\
  -\nbl_\+\Bigg[ \underbrace{\f{c_\rmS^2}{\Gamma} \f{\delta\! p}{p_0}}_{\epsilon^0\omega v_\rmA} + v_\rmA^2\Bigg( \underbrace{\f{1}{2}\f{|\delta\bm{B}|^2}{B_0^2}}_{\epsilon^1\omega v_\rmA} + \underbrace{\f{\delta\! B_\|}{B_0}}_{\epsilon^0\omega v_\rmA} \Bigg) \Bigg] -\f{\delta\rho}{\rho_0}\zhat\pp{}{z}\Bigg[ \underbrace{\f{c_\rmS^2}{\Gamma} \f{\delta\! p}{p_0}}_{\epsilon^2\omega v_\rmA} + v_\rmA^2\Bigg( \underbrace{\f{1}{2}\f{|\delta\bm{B}|^2}{B_0^2}}_{\epsilon^3\omega v_\rmA} + \underbrace{\f{\delta\! B_\|}{B_0}}_{\epsilon^2\omega v_\rmA} \Bigg) \Bigg]\\
  -\f{\delta\rho}{\rho_0}\nbl_\+\Bigg[ \underbrace{\f{c_\rmS^2}{\Gamma} \f{\delta\! p}{p_0}}_{\epsilon^1\omega v_\rmA} + v_\rmA^2\Bigg( \underbrace{\f{1}{2}\f{|\delta\bm{B}|^2}{B_0^2}}_{\epsilon^2\omega v_\rmA} + \underbrace{\f{\delta\! B_\|}{B_0}}_{\epsilon^1\omega v_\rmA} \Bigg) \Bigg] + v_\rmA^2\Bigg( \underbrace{\pp{}{z}\f{\delta\bm{B}}{B_0}}_{\epsilon^1\omega v_\rmA} + \underbrace{\f{\delta\! B_\|}{B_0}\pp{}{z}\f{\delta\bm{B}}{B_0}}_{\epsilon^2\omega v_\rmA} + \underbrace{\f{\delta\bm{B}_\+}{B_0}\cdot\nbl_\+\f{\delta\bm{B}}{B_0}}_{\epsilon^1\omega v_\rmA} \Bigg) \\
  - 2\Omega(-\underbrace{\cos\theta\,\yhat}_{\epsilon^2\omega v_\rmA} + \underbrace{\sin\theta\,\zhat}_{\epsilon^1\omega v_\rmA})\times\bm{u} + q\Omega u_x(\underbrace{\sin\theta\,\yhat}_{\epsilon^2\omega v_\rmA} + \underbrace{\cos\theta\,\zhat}_{\epsilon^1\omega v_\rmA}).
  \label{e:e.o.m. -2-}
\end{multline}
From the $\calO(\epsilon^0\omega v_\rmA)$ terms in \eqref{e:e.o.m. -2-}, one gets the pressure balance
\begin{align}
  \f{c_\rmS^2}{\Gamma} \f{\delta\! p}{p_0} + v_\rmA^2\f{\delta\! B_\|}{B_0} = 0,
  \label{e:e.o.m. O(0)}
\end{align}
which, when combined with \eqref{e:adiabatic}, becomes  
\begin{align}
  \f{\delta\! \rho}{\rho_0} + \f{v_\rmA^2}{c_\rmS^2} \f{\delta\! B_\|}{B_0} = 0.
  \label{e:e.o.m. O(0) -2-}
\end{align}
From the $\calO(\epsilon^1\omega v_\rmA)$ terms in \eqref{e:e.o.m. -2-}, we obtain
\begin{equation}
  \pp{\bm{u}}{t} + \bm{u}_\+\cdot\nbl_\+\bm{u} = - \nbl_\+\Bigg( \f{v_\rmA^2}{2}\f{|\delta\bm{B}|^2}{B_0^2} \Bigg) + v_\rmA^2\Bigg( \pp{}{z}\f{\delta\bm{B}}{B_0} + \f{\delta\bm{B}_\+}{B_0}\cdot\nbl_\+\f{\delta\bm{B}}{B_0} \Bigg) \\
   + 2\Omega\yhat\times\bm{u} + q\Omega u_x\zhat,
  \label{e:e.o.m. O(1)}
\end{equation}
where we have used $\cos\theta \simeq 1$ and neglected all terms containing $\sin \theta \ll 1$.
The desired perpendicular and parallel momentum equations \eqref{e:RMHD Phi} and \eqref{e:RMHD upar} are recovered as $\zhat\cdot[\nbl_\+\times$\eqref{e:e.o.m. O(1)}] and $\zhat\cdot$\eqref{e:e.o.m. O(1)}, respectively.

Next, the ordering of terms in \eqref{e:induction} is as follows:
\begin{multline}
  \underbrace{\pp{}{t}\f{\delta \bm{B}}{B_0}}_{\epsilon^1\omega} + \underbrace{u_\|\pp{}{z}\f{\delta \bm{B}}{B_0}}_{\epsilon^2\omega} + \underbrace{\bm{u}_\+\cdot\nbl_\+\f{\delta \bm{B}}{B_0}}_{\epsilon^1\omega} - \underbrace{q\Omega x\sin\theta\Bigg(\pp{}{y} +  \f{1}{\tan\theta}\pp{}{z}\Bigg)}_{\epsilon^2\omega}\f{\delta \bm{B}}{B_0} + \Bigg(\underbrace{\zhat}_{\epsilon^1\omega} + \underbrace{\f{\delta\bm{B}}{B_0}}_{\epsilon^2\omega}\Bigg)\pp{u_\|}{z} \\
  = \underbrace{\pp{\bm{u}}{z}}_{\epsilon^1\omega} + \underbrace{\f{\delta\! B_\|}{B_0}\pp{\bm{u}}{z}}_{\epsilon^2\omega} + \underbrace{\f{\delta\bm{B}_\+}{B_0}\cdot\nbl_\+\bm{u}}_{\epsilon^1\omega}  - q\Omega\f{\delta\! B_x}{B_0}( \underbrace{\sin\theta\yhat}_{\epsilon^2\omega} + \underbrace{\cos\theta\zhat}_{\epsilon^1\omega} ).
\end{multline}
Together with \eqref{e:continuity O(1)} and \eqref{e:e.o.m. O(0) -2-}, the $\calO(\epsilon^1\omega)$ terms in this equation yield
\begin{align}
  \lf( \pp{}{t} + \bm{u}_\+\cdot\nbl_\+ \ri)\lf( \f{\delta\bm{B}}{B_0} + \zhat\f{v_\rmA^2}{c_\rmS^2}\f{\delta\! B_\|}{B_0} \ri) = \lf( \pp{}{z} + \f{\delta\bm{B}}{B_0}\cdot\nbl_\+ \ri)\bm{u}  - q\Omega\f{\delta\! B_x}{B_0}\zhat.
  \label{e:induction O(1)}
\end{align}
Finally, we obtain the perpendicular and parallel magnetic field equations \eqref{e:RMHD Psi} and \eqref{e:RMHD Bpar} as $\zhat\cdot[\mr{curl}^{-1}$\eqref{e:induction O(1)}] and $\zhat\cdot$\eqref{e:induction O(1)}, respectively.

\section{Compressive-to-Alfv\'enic energy-injection ratio for a single linear MRI mode in RRMHD}\label{s:I_compr/I_AW}
Substituting the solution to the dispersion relation \eqref{e:RMHD dispersion} back into the linearized RRMHD equations \eqref{e:RMHD Psi}-\eqref{e:RMHD Bpar}, one gets the linear relations
\begin{subequations}
\begin{align}
  & \f{\delta\! B_{\|}}{B_0} = \lambda\f{k_y\Psi}{v_\rmA}, 
  \label{e:app2-1} \\
  & \Phi = -\f{\omega}{k_\| v_\rmA}\Psi, 
 \label{e:app2-2} \\
  & u_{\|} = -\f{\Omega}{k_\| v_\rmA}\lf[ \lf( 1 + \f{v_\rmA^2}{c_\rmS^2} \ri)\lambda\f{\omega}{\Omega} + q \ri]k_y\Psi,
 \label{e:app2-3} 
\end{align}
\end{subequations}
where, 
\begin{multline}
  \lambda = \f{5\rmi\beta}{2\sqrt{2}(k_y/k_\+)^2(6 + 5\beta)^{3/2}}\Big[ -6\bar{k}_\|^2 + 7(k_y/k_\+)^2(6 + 5\beta) \\
  - \sqrt{36\bar{k}_\|^4 + (k_y/k_\+)^2(6 + 5\beta)^2 + 4(k_y/k_\+)^2\bar{k}_\|^2(6 + 5\beta)(3 + 20\beta)} \Big]\Big[\bar{k}_\|^2(6 + 10\beta) + (k_y/k_\+)^2(6 + 5\beta) \\
  - \sqrt{36\bar{k}_\|^4 + (k_y/k_\+)^2(6 + 5\beta)^2 + 4(k_y/k_\+)^2\bar{k}_\|^2(6 + 5\beta)(3 + 20\beta)}\Big]^{-1},
\end{multline}
with $\bar{k}_\| = k_\|v_\rmA/\Omega$.
For the fastest-growing mode, $\lambda$ reduces to $\sqrt{-5\beta/(5\beta + 6)}$.
Substituting \eqref{e:app2-1}-\eqref{e:app2-3} into \eqref{e:I_AW} and \eqref{e:I_compr}, one obtains 
\begin{equation}
  \f{I_\mr{compr}}{I_\mr{AW}} = 
  -\f{ q\lambda( k_\| v_\rmA )^2/[ ( 1 + v_\rmA^2/c_\rmS^2 )\lambda\Omega/\omega + q] + (2 - q)\Omega\omega^*}{2\Omega\omega^*},
  \label{e:I_compr/I_AW}
\end{equation}
where the superscript star denotes the complex conjugate.
Note that, when the rotation is not sheared, i.e., $q = 0$, this becomes the conservation of energy $I_\mr{compr} + I_\mr{AW} = 0$, i.e., the Alfv\'enic and compressive fluctuations exchange their energy via unsheared rotation.


\bibliographystyle{jpp}
\bibliography{references}

\end{document}